\documentclass[aps,prx,twocolumn,showpacs,superscriptaddress,longbibliography]{revtex4-1}

\usepackage{bbm}
\usepackage{amsmath}
\usepackage{hyperref}
\usepackage{graphicx}
\usepackage{graphics}
\usepackage{amssymb}
\usepackage{mathtools}
\usepackage{xcolor}
\usepackage{physics}
\usepackage[british]{babel}
\usepackage{csquotes}
\usepackage{graphicx}

\graphicspath{{Figs/}}

\begin{document}

\title{Investigating Spectral Dynamics and Spin Signatures of a Mechanically Isolated Quantum Emitter in hBN}

\author{Sajedeh Shahbazi}
\thanks{These authors contributed equally to this work}
\author{Alexander Pachl}
\thanks{These authors contributed equally to this work}
\author{Kathrin Schwer}
\author{Patrick Maier}
\author{Alexander Kubanek}
\email{alexander.kubanek@uni-ulm.de}
\affiliation{Institute for Quantum Optics, Ulm University, D-89081 Ulm, Germany}


\begin{abstract}
Mechanically isolated defect centers in hexagonal boron nitride are promising coherent quantum emitters, yet spectral instabilities persist, and their spin-related nature remains unclear. Here we investigate a single mechanically isolated quantum emitter in hBN integrated onto a coplanar waveguide. The emitter exhibits exceptionally bright photoluminescence excitation signals with saturation count rates exceeding $10\,\mathrm{Mc/s}$. High-resolution spectroscopy reveals two closely spaced zero-phonon-line transitions originating from the same defect complex. Time-resolved spectroscopy shows that these transitions exhibit markedly different spectral diffusion dynamics, consistent with distinct donor–acceptor–pair–like recombination pathways with different sensitivities to local electrostatic fluctuations. Off-resonant blue illumination redistributes emission between the two transitions and increases the emission duty cycle without significantly modifying the dominant spectral diffusion rates at low temperature, indicating repumping from long-lived shelving states. Magnetic-field-dependent photoluminescence, optically detected magnetic resonance, and pump–probe measurements reveal millisecond-scale relaxation dynamics and magnetic-field-dependent fluorescence contrast, demonstrating spin-dependent population dynamics in the metastable shelving state. These results clarify how charge-driven spectral fluctuations and spin-dependent shelving jointly shape the optical cycling dynamics.

\end{abstract}

\maketitle

\section{Introduction}
Defect centers in hexagonal boron nitride (hBN) have emerged as bright and spectrally narrow solid-state quantum emitters compatible with van der Waals material platforms~\cite{bourrellier_bright_2016,dietrich_observation_2018,grosso_tunable_2017}. A variety of defect families have been reported in hBN, showing a wide range of characteristics and applications. Among these, mechanically isolated quantum emitters are attractive due to their narrow optical linewidths, which can persist up to room temperature~\cite{dietrich_solid-state_2020}. Their reduced coupling to low-energy in-plane phonons makes them promising candidates for coherent optical control. Indeed, resonant optical driving and Rabi oscillations have been demonstrated at cryogenic temperatures up to $20\,\mathrm{K}$~\cite{koch_probing_2024}. Despite these encouraging results, key questions remain regarding the microscopic origin of these emitters, the mechanisms underlying their spectral instabilities, and the existence of optically accessible spin states.

Several defect families have been proposed in hBN and demonstrated to host optically addressable spin states~\cite{exarhos_magnetic-field-dependent_2019,gottscholl_room_2021,stern_quantum_2024,auburger_towards_2021}. The coexistence of bright, coherent optical transitions and microwave-addressable spin degrees of freedom raises the prospect of spin–photon interfaces; nevertheless, spectral instability remains a central challenge for coherent optical control.

Emitters frequently exhibit spectral diffusion, discrete spectral jumps, and fluorescence intermittency (blinking) on timescales ranging from nanoseconds to seconds~\cite{white_phonon_2021,grosso_tunable_2017,boll_photophysics_2020}. Spectral diffusion refers to stochastic temporal fluctuations of the optical transition energy, typically caused by changes in the local electrostatic, dielectric, or strain environment. Depending on the microscopic origin and the relevant observation timescale, these fluctuations may appear as continuous spectral wandering or as abrupt switching between distinct emission energies. Blinking and long-lived shelving states further interrupt optical cycling and reduce the emission duty cycle. These effects are commonly attributed to charge trapping and detrapping, local electric-field fluctuations, and coupling to nearby defects~\cite{galland_two_2011,akbari_temperature-dependent_2021,li_quantum_2025}. Such instabilities broaden the optical linewidth, reduce photon indistinguishability, and complicate resonant coherent driving, requiring that the optical Rabi frequency exceed the characteristic spectral diffusion rate.

Recent studies interpret several hBN emitters within a donor–acceptor-pair (DAP) framework, in which radiative recombination occurs between spatially separated donor and acceptor states~\cite{museur_defect-related_2008,tan_donoracceptor_2022}. In such a picture, the emission energy depends sensitively on the donor–acceptor separation and the local electrostatic environment, making DAP transitions naturally susceptible to spectral diffusion~\cite{li_quantum_2025}. Various recombination pathways within a single defect complex can therefore lead to multiple zero-phonon lines (ZPLs) with different sensitivities to environmental charge fluctuations~\cite{mejia_dynamic_2025}. However, whether such pathway-dependent recombination channels exhibit distinct spectral dynamics and how they relate to long-lived shelving processes remains largely unexplored.

Spin-based measurements introduce complementary constraints. Optically detected magnetic resonance (ODMR) and pulsed pump–probe protocols rely on spin-selective optical cycling and population transfer into metastable configurations. The efficiency of spin initialization and readout depends not only on the spin Hamiltonian but also on branching ratios between bright and dark states, which can vary with excitation conditions and magnetic-field orientation~\cite{udvarhelyi_planar_2023}. A unified understanding of spectral diffusion, blinking, and spin-dependent shelving is therefore essential for optimizing both optical coherence and spin control.

In this work, we investigate a mechanically isolated single emitter in hBN that exhibits high brightness under resonant excitation. Within the donor–acceptor pair (DAP) framework, we examine its spectral stability. High-resolution spectroscopy reveals two closely spaced ZPL transitions originating from the same emitter that exhibit distinct spectral diffusion dynamics, suggesting separate recombination pathways with different sensitivities to local electrostatic fluctuations. We further investigate the emitter's spin dynamics using magnetic-field-dependent photoluminescence, ODMR, and time-resolved pump–probe measurements. These measurements reveal long-lived relaxation dynamics on the millisecond timescale and excitation-dependent population pathways.

\section{Methods}
\subsection*{Sample preparation and setup}
The hBN samples were prepared by dissolving commercially available micro-sized hBN powder (2D Semiconductors) in ethanol and spin-coating the solution onto a tapered coplanar waveguide. After deposition, the sample was annealed under vacuum at $800\,^\circ\mathrm{C}$ for 1 hour. The sample was cooled from $800\,^\circ\mathrm{C}$ to room temperature over a period of 2 hours.

The coplanar waveguide was fabricated on a sapphire substrate using standard optical lithography and lift-off processes. It consists of a $200\,\mathrm{nm}$-thick gold layer deposited on a $20\,\mathrm{nm}$ titanium adhesion layer. The width of the center conductor and the gap spacing were chosen as $47\,\mu\mathrm{m}$ and $20\,\mu\mathrm{m}$, respectively, to achieve an impedance of $50\,\Omega$ and minimize microwave reflections. The hBN emitter studied in this work is located on the signal line in proximity to the edge of the gap where the microwave electric field is strongest.

The sample was then mounted on a custom copper cold finger and placed in a vacuum-sealed flow cryostat (CryoVac) that can be cooled with liquid nitrogen or liquid helium.
A high-NA (NA=0.9) objective was mounted inside the cryostat, enabling efficient excitation and emission of the hBN defect with a confocal setup. For acquiring PL spectra, an off-resonant $532\,\mathrm{nm}$ laser was used to excite the emitter, and the emission was collected by the spectrometer after filtering the excitation laser with a 550 nm long-pass filter. For PLE measurements, a tunable Dye laser was used to excite the emitter resonantly and the signal from the phonon sideband was collected after filtering the ZPL signal with a tunable long-pass filter. The optical power from both off-resonant and resonant lasers was stabilized using a home-built PID controller feedback loop connected to an acousto-optic modulator (AOM), resulting in relative power fluctuations below 1\% over the investigated power range. Photon detection was performed using avalanche photodiodes (SPCM-AQRH-14-FC, Excelitas), which exhibit a photon detection efficiency exceeding 70\% at a wavelength of $650\,\mathrm{nm}$.

Magneto-photoluminescence measurements were performed using a horseshoe permanent magnet mounted on a motorized rotation stage positioned in close proximity to the sample. The magnet was aligned such that the magnetic field was predominantly in the plane of the sample surface. Based on magnetostatic simulations and considering the uncertainty in the emitter position over the lateral scale of the sample, we estimate a possible deviation of the field direction from the sample plane of approximately $\pm 12^\circ$. To estimate the magnetic-field amplitude, we additionally placed NV nanocrystals on the coplanar waveguide and measured their ODMR response in the presence of the applied magnetic field. Combining these ODMR measurements with magnetostatic simulations and an estimated magnet--sample separation of $6 \pm 1 \,\mathrm{mm}$, we estimate the magnetic field at the emitter location to be approximately $40 \pm 10 \,\mathrm{mT}$.

\section{ultra bright mechanically isolated  emitter}

\begin{figure*}
\hspace{0cm}{\includegraphics[scale=1]{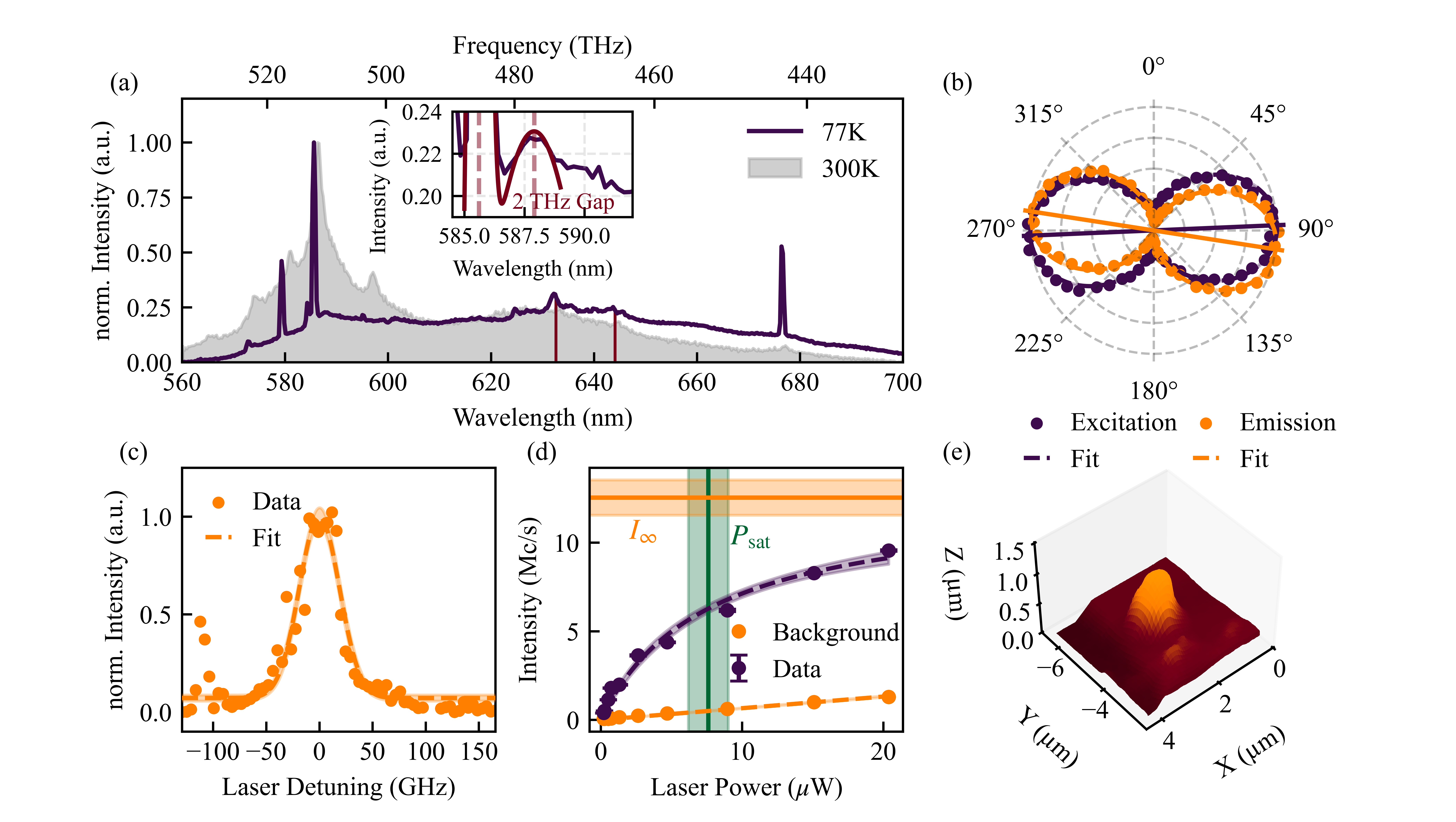}} 
\caption{Photophysical characterization of a single emitter in hBN. (a) Photoluminescence spectrum of the emitter recorded at 77~K (purple). The inset shows a zoom-in of the ZPL region, highlighting the spectral gap between the ZPL and the first acoustic phonon sideband of $2 \pm 0.5\,\mathrm{THz}$. The dashed lines mark the centers of Gaussian fits (dark red) to both peaks, from which the gap is extracted. The two phonon sidebands marked by red lines at $631.6\,\mathrm{nm}$ and $645.0\,\mathrm{nm}$ are characteristic signatures of mechanically isolated hBN emitters. The gray-shaded region shows the PL spectra of the emitter at room temperature. 
(b) Polarization-resolved photoluminescence for both excitation (purple) and emission (orange) under off-resonant excitation, detected through a narrow band-pass filter centered on the ZPL. 
(c) Photoluminescence excitation scan across the ZPL using a tunable resonant laser. The data are fitted with a Gaussian (dashed line), yielding an inhomogeneous linewidth of $44 \pm 3\,\mathrm{GHz}$.
(d) Saturation behavior (purple) extracted from time-resolved fluorescence traces acquired at different resonant excitation powers after subtraction of the linearly increasing background contribution (orange). (e) AFM scan of the emitter located on the gold-coated region of the coplanar waveguide.
}
\label{fig1}
\end{figure*}

Figure~\ref{fig1}(a) shows the photoluminescence (PL) spectrum of the emitter measured at $77\,\mathrm{K}$, exhibiting a zero-phonon line (ZPL) at $585\,\mathrm{nm}$ with linearly polarized emission, as shown in Fig.~\ref{fig1}(b). A weaker emission line is also observed at $579\,\mathrm{nm}$ with a different polarization dipole indicating that this feature originates from a neighboring emitter located close to the investigated emitter and is therefore not associated with the defect studied in this work. The PL spectrum further shows the presence of an energy gap between the ZPL and the first acoustic phonon mode, with a value of $2 \pm 0.5\,\mathrm{THz}$ (inset of Fig.~\ref{fig1}(a)). This gap suggests a reduced coupling between the electronic orbital states and low-energy phonon modes. The presence of two optical phonon sidebands at $631.6\,\mathrm{nm}$ and $645.0\,\mathrm{nm}$, corresponding to the phonon energies of $157$ and $197\,\mathrm{meV}$, respectively, is consistent with previously reported mechanically isolated emitters in hexagonal boron nitride~\cite{hoese_mechanical_2020,dietrich_observation_2018,hoese_single_2022}.

Fig.~\ref{fig1}(c) shows a photoluminescence excitation (PLE) scan of the ZPL using a tunable resonant laser. The emission peak exhibits spectral diffusion; consequently, temporal fluctuations in its spectral position result in spectral line broadening. The distribution of peak positions extracted from multiple scans follows a Gaussian profile with a full width at half maximum of $44 \pm 3\,\mathrm{GHz}$, corresponding to the inhomogeneous linewidth. 
After performing the PLE measurements, resonant saturation measurements were carried out by setting the laser wavelength at the center of the inhomogeneous linewidth. Time traces of the emitted fluorescence were then recorded using a single-photon counter module. The bin size for the recorded time traces was chosen based on the spectral diffusion timescale of the emitter, which is discussed in detail in the following section. By recording time traces at different resonant laser powers, the emitter's saturation behavior was extracted. An example of a recorded time trace is shown in the Appendix~A, Fig.~\ref{fig:photon statistics}(a) and (b). For each excitation power, the emitter was classified as being in the ON state whenever the detected count rate exceeded the threshold indicated in Fig.~\ref{fig:photon statistics}(b). The saturation curve was then constructed from the average photon count rate of multiple ON-state intervals.

The measured count rate as a function of excitation power was fitted using the standard saturation model,
\begin{equation}
I(P) = I_{\infty}\,\frac{P}{P + P_{\mathrm{sat}}},
\end{equation}
where \(I(P)\) is the detected fluorescence count rate at excitation power \(P\), \(I_{\infty}\) is the saturation count rate, and \(P_{\mathrm{sat}}\) is the saturation power (Fig.~\ref{fig1}(d)). From the fit, we obtain a saturation power of \(P_{\mathrm{sat}} = 7.6 \pm 1.5\,\mu\mathrm{W}\) and a saturation count rate of \(I_{\infty} = 12.5 \pm 1.5\,\mathrm{Mc/s}\). A weak background contribution exhibiting an approximately linear dependence on the excitation power was observed and excluded from the analysis prior to fitting. Consequently, the saturation model was applied only to the emitter fluorescence after background subtraction. The high saturation count rate can be analyzed considering the detection efficiency of the setup, the Debye–Waller factor, and the quantum efficiency of the emitter. For a microscope objective with a numerical aperture of 0.9, the collection efficiency for an in-plane electric dipole in a homogeneous medium is approximately $33~\%$. This value represents a lower bound, since it is calculated assuming an air interface. In the present experiment, the emitter is located on a gold substrate, where the angular emission pattern can be significantly modified. Fig.~\ref{fig1}(e) shows the AFM image of the emitter located within a micrometer-scale cluster of hBN flake. Depending on the emitter–surface separation, interference with the reflected field and modifications of the local density of optical states, including plasmon-mediated effects, can preferentially redirect emission toward the air side. This redirection can increase the fraction of photons collected by the objective and lead to more directional emission~\cite{Mattheyses2005}.
Taking into account the objective collection efficiency, the transmission of the detection path, and the detector quantum efficiency, the overall detection efficiency of the setup is estimated to be approximately $10\%$, without considering possible modifications due to the gold substrate.

From second-order autocorrelation measurements, an excited-state lifetime of $\tau = 1.26 \pm 0.11 \, \mathrm{ns}$ is extracted, corresponding to a 
Fourier-transform-limited linewidth of 
$ 1/(2\pi\tau) = 126 \pm 11\,\mathrm{MHz}$. The corresponding maximum decay rate, $\Gamma_{\mathrm{max}} = 1/2\tau$, is $\mathrm(397 \pm 35)\times10^{6}\,\mathrm{s^{-1}}$. The Debye--Waller factor of the emitter is estimated to be approximately $20~\%$. To determine this value, the ZPL contribution was obtained from the integrated area of a Gaussian fit to the ZPL peak in Fig.~\ref{fig1}(a). The phonon-sideband contribution was independently determined by resonantly exciting the emitter and detecting its phonon-sideband emission, thereby strongly suppressing the broadband background contribution. Since the off-resonant and resonant spectra were acquired under different excitation conditions, the resonant excitation was rescaled to the off-resonant PL spectrum by matching the amplitude of the optical phonon sideband, such that the PSB maxima in the two spectra overlap. 
Taking into account the measured saturation count rate under resonant excitation, the detection efficiency and the phonon sideband fraction, the lower limit for the quantum efficiency of this emitter is estimated to be $33 \pm 9 ~\%$.
Direct measurements of the quantum efficiency of single-photon emitters in hBN have reported values ranging from $40~\%$ up to a record value of $87~\%$ for emitters with zero-phonon lines near  $580\,\mathrm{nm}$ ~\cite{Nikolay:19, Castelletto2020}, placing the value obtained here well within the range reported in the literature.
 
\section{ZPL spectral structure and dynamics of DAP-like pathways}

\begin{figure*}[!htbp]
\hspace{0cm}{\includegraphics[scale=0.9]{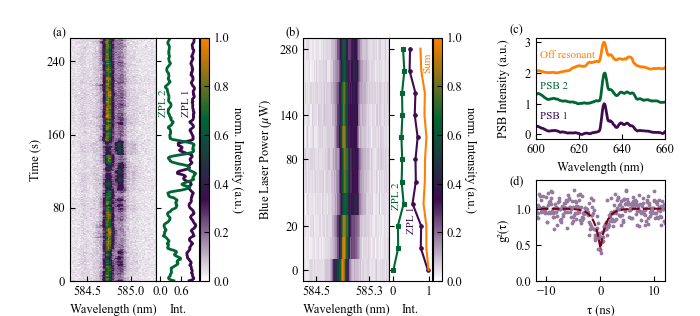}} 
\caption{Correlations between two closely spaced ZPL transitions from a single emitter. (a) Time-resolved photoluminescence spectra under off-resonant green excitation, showing discrete spectral switching between two closely-spaced ZPL transitions, with $\mathrm{ZPL}_1$ (purple) as the dominant peak and $\mathrm{ZPL}_2$ (green) as the weaker emission line.
(b) Photoluminescence spectra illustrating the effect of additional off-resonant blue illumination, which redistributes the emission intensity between the two ZPL peaks. The corresponding ZPL intensities on the right panel were extracted from Gaussian fits to the two individual ZPL peaks, with the orange curve indicating the sum of the two fitted Gaussians.
(c) Phonon sideband spectra obtained under resonant excitation of each ZPL transition (purple and green), showing identical sideband profiles that also coincide with the phonon sideband observed under off-resonant excitation (orange spectrum).The spectra are vertically offset by +1 for clarity.
(d) Second-order autocorrelation function measured under off-resonant green excitation, showing $g^{(2)}(0)=0.443$.
}
\label{secondpeak with blue}
\end{figure*}

A high-resolution PL spectrum reveals two closely spaced ZPL lines separated by approximately $\Delta \lambda = 0.12\,\mathrm{nm}$. The intensity of the weaker line is approximately one-third of the dominant peak. This observation raises the question of whether the two lines originate from the same emitter or from two distinct emitters.

Off-resonant, second-order autocorrelation measurements, taking both ZPLs into account, yield $g^{(2)}(0) = 0.443$, indicating primarily single-photon emission from a single emitter [Fig.\ref{secondpeak with blue}(d)]. However, since the second ZPL line is weaker than the dominant peak, its contribution to the measured autocorrelation function is reduced, and this measurement alone cannot conclusively rule out the presence of two closely spaced emitters.

To further investigate the origin of the two ZPL lines, time-resolved PL spectroscopy was performed. Figure~\ref{secondpeak with blue}(a) shows a sequence of PL spectra recorded under off-resonant green excitation at $532\,\mathrm{nm}$. Each spectrum was acquired with an integration time of $1\,\mathrm{s}$ over a total duration of $250\,\mathrm{s}$. The intensities of the two ZPL peaks fluctuate with time. Importantly, these fluctuations are strongly anticorrelated: when the intensity of the dominant peak decreases, the intensity of the weaker peak increases.

Further insight is obtained by introducing an additional off-resonant blue laser at a wavelength of $405\,\mathrm{nm}$, as shown in Fig.~\ref{secondpeak with blue}(b). As blue laser power increases, the intensity of the dominant ZPL peak decreases, while that of the weaker peak increases. This redistribution of intensity again exhibits a clear anticorrelation, with the total integrated ZPL intensity remaining constant. These observations indicate that the blue excitation modifies the population branching between two emissive states of the same emitter rather than activating a second independent emitter.

Additional confirmation is provided by resonant excitation measurements. The phonon sideband spectra obtained by resonantly exciting each of the two ZPL peaks overlap and are identical to the sideband observed under off-resonant excitation, as shown in Fig.~\ref{secondpeak with blue}(c). The identical phonon sideband profiles indicate that both ZPL transitions couple to the same type of phonon modes.

The observed anticorrelated intensity fluctuations between the two closely spaced ZPLs, together with the conservation of their total integrated intensity and identical phonon sideband spectra, indicate that the two lines correspond to competing radiative channels of a single defect complex. This behavior is consistent with recent studies that attribute spectral jump and line switching in hBN to the dynamical selection of nonlocal donor--acceptor-pair (DAP)-like recombination pathways involving electronic states derived from nitrogen $\pi^{*}$ orbitals~\cite{mejia_dynamic_2025, pelliciari_elementary_2024}. The introduction of off-resonant blue illumination further modifies the relative intensities of the two ZPL transitions while conserving their total emission, indicating that the blue laser perturbs the local charge configuration and alters the branching ratio between these competing recombination pathways.

\section{Power-Dependent Spectral Dynamics Under Resonant Excitation at 77~K and 8~K}

Spectral diffusion is a major limitation for emitters in hBN, as it can hinder coherent optical control. In this section, we investigate the spectral diffusion dynamics of the two closely-spaced ZPL transitions under resonant excitation as a function of excitation power, at two different temperatures, and in the presence of additional off-resonant laser excitation.

Spectral diffusion measurements were performed by tuning the resonant laser to the center of the inhomogeneous ZPL linewidth and recording the fluctuations of the time-resolved fluorescence traces. The emitter was classified in the ON state when the detected count rate exceeded a threshold of $3\sigma$, where $\sigma$ denotes the standard deviation of the background count rate; all lower count rates were assigned to the OFF state. An example time trace together with the applied threshold is shown in Appendix~A, Fig.~\ref{fig:photon statistics}(a),(b).

\subsection{Spectral diffusion of the dominant ZPL transition}

\begin{figure}[!htbp]
\hspace{0cm}{\includegraphics[scale=1]{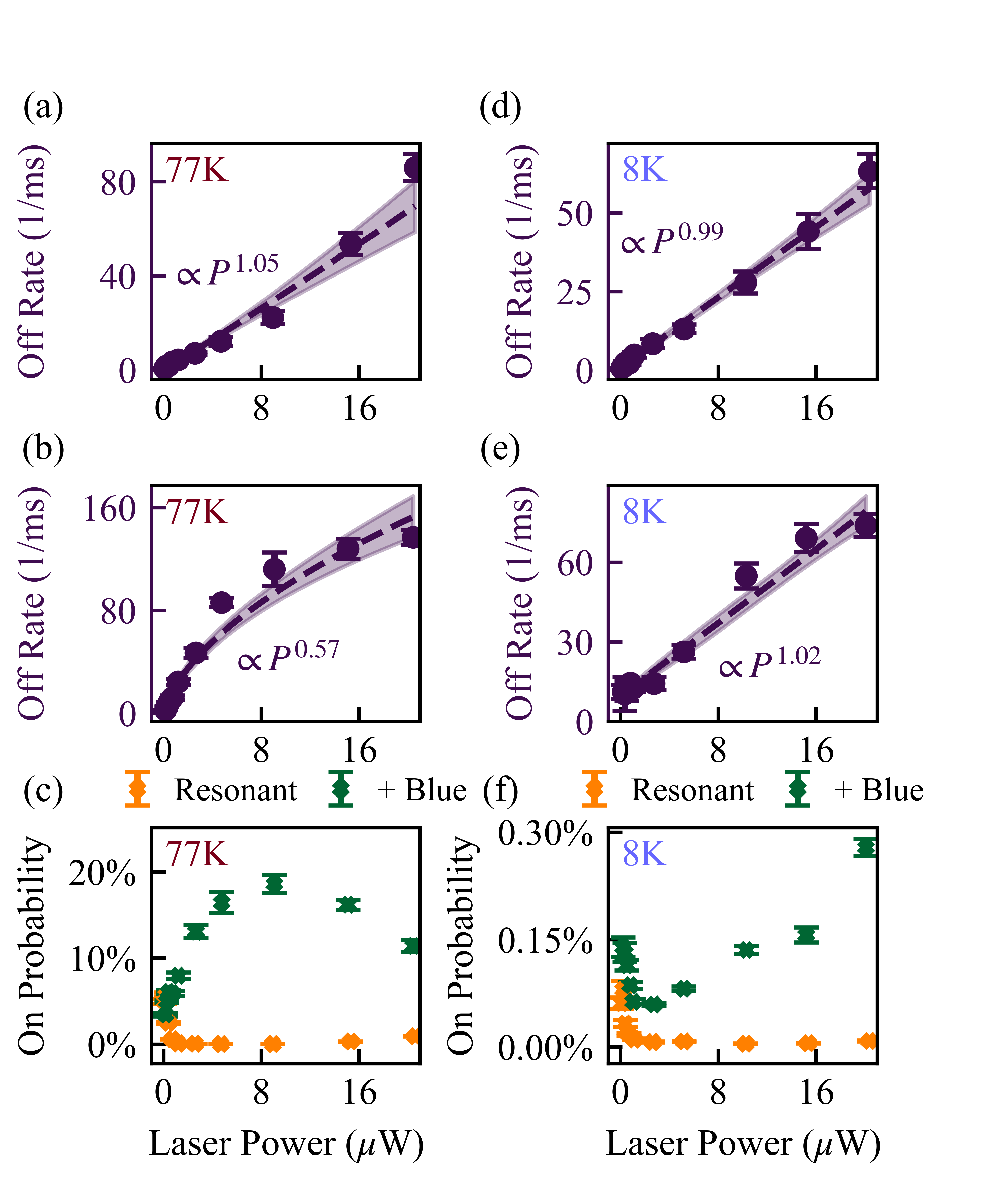}}
\caption{Spectral dynamics of the dominant ZPL transition. (a,b) Power-dependence of the off-rate (purple) of the dominant ZPL transition measured at 77~K under resonant excitation only (a) and with the addition of a $5\,\mathrm{\mu\text{W}}$ blue laser (b). 
(c) Corresponding power dependence of the on-state probability at 77~K, comparing resonant excitation with (green) and without (orange) blue illumination. 
(d,e) Power dependence of the off-rate of the dominant ZPL transition measured at 8~K under resonant excitation only (d) and with additional blue illumination (e). 
(f) Corresponding power dependence of the on-state probability at 8~K.
}
\label{first peak SD}
\end{figure}

Figure~\ref{first peak SD} summarizes the spectral diffusion dynamics of the dominant ZPL transition under resonant excitation at 77~K and 8~K. For each resonant excitation power, independent time traces were recorded and analyzed to extract the durations of ON and OFF intervals. The resulting distributions follow approximately exponential decays, allowing characteristic time constants to be obtained from linear fits to semi-logarithmic histograms. The off-rate, which quantifies how rapidly the emitter switches to the off state, is given by the inverse of the average ON time.

At 77~K, the off-rate increases monotonically with the resonant excitation power [Fig.~\ref{first peak SD}(a)], indicating that resonant driving accelerates the spectral diffusion dynamics. When an off-resonant blue laser is introduced [Fig.~\ref{first peak SD}(b)], the off-rate increases by approximately a factor of two across the explored power range and the scaling becomes sub-linear. Although the average ON time is reduced under additional blue illumination, the probability of finding the emitter in the ON state, i.e., the fraction of time the emitter is active, increases from below 1\% to nearly 20\% for certain excitation powers [Fig.~\ref{first peak SD}(c)]. This behavior indicates that blue illumination modifies the switching kinetics by activating additional charge-related processes in the local environment, while biasing the system toward the resonant configuration.

Upon cooling to 8~K [Fig.~\ref{first peak SD}(d)], the spectral diffusion rates decrease only moderately. For example, at a resonant power of $20~\mu$W, the off-rate decreases from $\sim 85$~kHz at 77~K to $\sim 63$~kHz at 8~K, indicating weak temperature dependence of the dominant noise source mechanism. At this temperature, the addition of blue illumination has little effect on the extracted off-rate, and the power dependence becomes approximately linear [Fig.~\ref{first peak SD}(e)], demonstrating that thermally activated charge dynamics are largely frozen out. Nevertheless, blue illumination still increases the ON-state probability [Fig.~\ref{first peak SD}(f)], consistent with repumping from a long-lived shelving state—potentially involving a spin-related configuration—into the radiative manifold. We also performed saturation measurements with the blue laser switched on, for which the extracted saturation count rate remained within the error bars of the value obtained without blue illumination. This is consistent with the interpretation that the blue laser primarily increases the duty cycle and ON-state probability, rather than the fluorescence count rate during the ON state.

\subsection{Spectral diffusion of the non-dominant ZPL transition}

\begin{figure}[!htbp]
\hspace{0cm}{\includegraphics[scale=1]{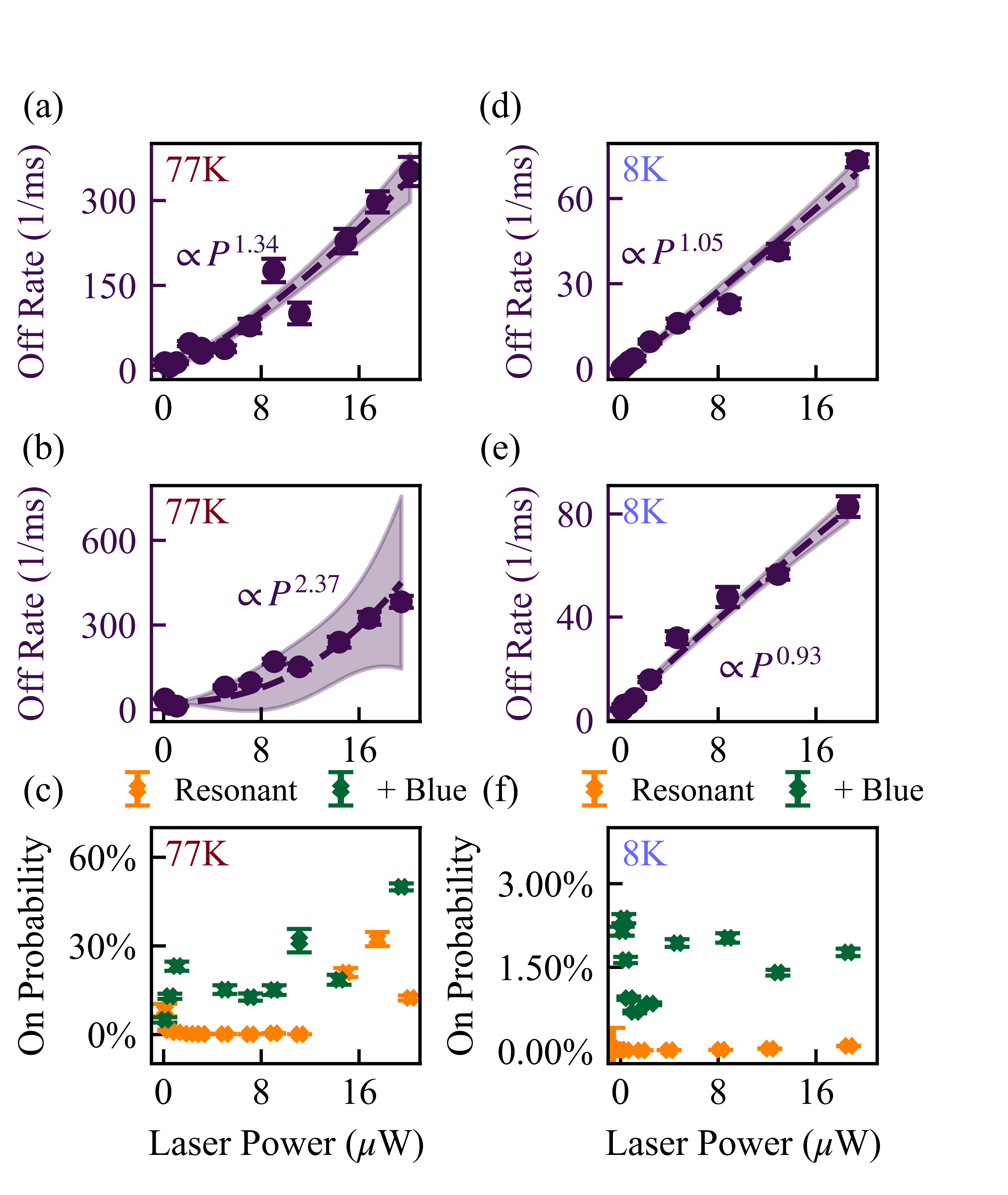}} 
\caption{Spectral dynamics of the non-dominant ZPL transition. (a,b) Power-dependent switching rates of the non-dominant ZPL transition at 77~K, showing the off-rate (purple) under resonant excitation only (a) and in the presence of additional off-resonant blue illumination (b). 
(c) On-state probability as a function of resonant excitation power at 77~K, comparing measurements with (green) and without (orange) blue illumination. 
(d,e) Power dependence of the off-rates of the non-dominant ZPL transition at 8~K under purely resonant excitation (d) and with added blue illumination (e). 
(f) Corresponding power dependence of On-state probability at 8~K.
}
\label{second peak SD}
\end{figure}

An analogous analysis was performed for the non-dominant ZPL transition, as shown in Fig.~\ref{second peak SD}. At 77~K, this transition exhibits substantially stronger spectral diffusion than the dominant peak, with off-rates approximately three times larger over the explored excitation powers. In contrast to the dominant transition, the introduction of blue illumination has only a minor effect on the extracted switching rates, indicating that the diffusion mechanism for this transition is largely insensitive to photo-induced charge reconfiguration.

A pronounced temperature dependence is observed for the non-dominant transition. Upon cooling from 77~K to 8~K, the off-rate is reduced by approximately a factor of five, demonstrating that the dominant diffusion mechanism is strongly thermally activated. Despite this strong temperature dependence, blue illumination does not significantly alter the spectral diffusion rates at either temperature, while consistently increasing the ON-state probability. This indicates that off-resonant excitation primarily affects population dynamics, such as repumping from long-lived shelving states, rather than the fluctuator responsible for spectral diffusion.

The off-rate of the non-dominant transition exhibits a non-linear power dependence under blue illumination at 77~K, suggesting that multiple thermally activated dynamical processes contribute to the switching behavior in this regime. At 8~K, the power dependence becomes approximately linear, consistent with the suppression of thermally assisted charge- or configurational-pathway processes.

\subsection{Distinct spectral dynamics arising from separate DAP-like recombination pathways}

The contrasting spectral diffusion behavior of the dominant and non-dominant ZPL transitions provides insight into the microscopic origin of the emission dynamics. Despite originating from the same defect complex, the two transitions exhibit different sensitivities to temperature and off-resonant excitation. The dominant transition displays relatively weak spectral diffusion with modest temperature dependence, whereas the non-dominant transition exhibits larger diffusion, which is strongly suppressed at low temperature, indicating coupling to a thermally activated fluctuator.

These observations suggest that the two optical transitions correspond to distinct radiative channels that differ in their effective charge distribution and electrostatic sensitivity. Although both channels experience the same local lattice and vibrational environment, differences in their electronic configuration lead to distinct coupling strengths to environmental fluctuations. The response to blue illumination further supports this interpretation, revealing that population dynamics and spectral diffusion can be tuned independently.

\section{Magnetic-field-dependent spin photodynamics and relaxation}

To investigate the presence of a spin-dependent optical response, magneto-photoluminescence measurements were performed. The PL signal was spectrally filtered with a band-pass filter transmitting wavelengths between $583\,\mathrm{nm}$ and $595\,\mathrm{nm}$ to isolate the ZPL emission. Under off-resonant green excitation, the magnet was rotated within the sample plane while recording the integrated ZPL photon counts as a function of magnetic-field orientation.

\begin{figure}[!htbp]
\hspace{0cm}{\includegraphics[scale=1]{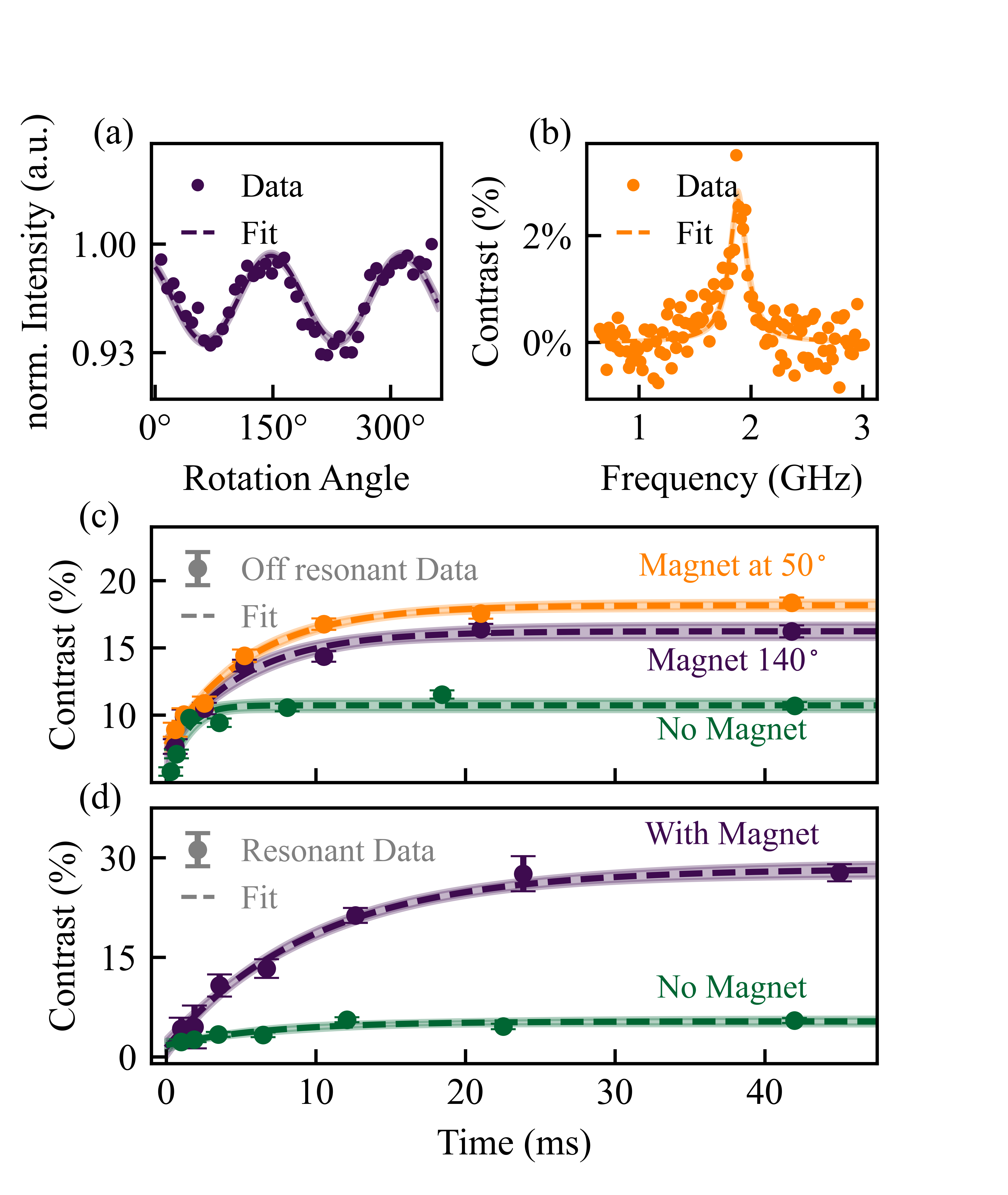}} 
\caption{Magnetic-field-dependent photoluminescence and spin relaxation of the emitter. (a) Integrated zero-phonon line photoluminescence intensity as a function of magnetic-field angle under off-resonant excitation. A sinusoidal modulation is observed, with a minimum near $50^\circ$ and a maximum near $140^\circ$, indicating magnetic-field-dependent spin mixing. (b) ODMR spectrum measured at $77~\mathrm{K}$ in the presence of an external magnetic field, showing a resonance at $(1.87 \pm 0.10)\,\mathrm{GHz}$ with a contrast of $(2.65 \pm 0.20)\%$.
(c) Pump--probe recovery measurements under off-resonant excitation at $77~\mathrm{K}$. Orange circles correspond to a magnetic-field orientation of $50^\circ$, purple circles to $140^\circ$, and green circles to zero magnetic field. Dashed lines are single-exponential fits used to extract the relaxation time $T_1$.(d) Pump--probe recovery measurements under resonant excitation. Purple circles denote measurements in the presence of the magnetic field, and green circles denote measurements without the magnetic field. Dashed lines are single-exponential fits. The magnetic-field dependence of both the recovery time and fluorescence contrast indicates spin-dependent relaxation dynamics.
}
\label{spin part}
\end{figure}

Figure~\ref{spin part}(a) shows the ZPL-integrated PL intensity as a function of magnetic-field angle. A sinusoidal modulation of the emission intensity is observed, with a minimum near $50^\circ$ and a maximum near $140^\circ$. Such angular dependence is consistent with magnetic-field-induced modification of spin populations through anisotropic spin mixing, as previously reported for hBN quantum emitters~\cite{exarhos_magnetic-field-dependent_2019}. The observed modulation indicates that the emitter's optical cycling efficiency is influenced by the orientation of the external magnetic field, suggesting the presence of a spin-dependent relaxation pathway.

To directly probe microwave-addressable spin transitions, ODMR measurements were performed at $77\,~\mathrm{K}$ in the presence of the external magnetic field. A resonance was observed at $(1.87 \pm 0.10)\,\mathrm{GHz}$ with a contrast of $(2.65 \pm 0.20)\%$(Fig.~\ref{spin part}(b)). The ODMR spectrum was recorded for both $140^\circ$ and $50^\circ$ magnet orientations; however, no significant variation in contrast or linewidth was observed between the two angles within experimental uncertainty.

An ODMR resonance near $1.87\,\mathrm{GHz}$ has been reported for carbon-related spin defects in hBN exhibiting similar PL spectra~\cite{stern_quantum_2024,whitefield_narrowband_2026}. The measured full width at half maximum (FWHM) of the resonance in our experiment was approximately $200\,\mathrm{MHz}$. This linewidth is substantially broader than intrinsic values reported for isolated single defects under optimized conditions and may originate from unresolved hyperfine structure or the overlap of two nearby spin transitions~\cite{chejanovsky_single-spin_2021,gao_single_2025,gottscholl_room_2021}. At an applied field of approximately $40\,\mathrm{mT}$, misalignment between the static field, microwave driving field, and defect quantization axis can further contribute to inhomogeneous broadening~\cite{barry_sensitivity_2020}.

The ODMR signal was reproduced multiple times within the same cooling cycle and was spatially localized to the emitter position in the confocal scan. However, the resonance was not reproducible after a subsequent thermal cycle. Simultaneous measurements of the ZPL position revealed spectral shifts between $584.7\,\mathrm{nm}$ and $585.5\,\mathrm{nm}$ [Fig.~\ref{appendix magnetic part}(c)] across different cooldowns. This corresponds to an energy shift of approximately $3\,\mathrm{meV}$, within the range previously reported for hBN defects and attributed to strain-induced modifications of the local crystal field as well as electrostatic fluctuations giving rise to Stark shifts and spectral diffusion~\cite{grosso_tunable_2017,noh_stark_2018, white_phonon_2021}. In hBN defects, variations in the local crystal field (e.g., due to strain or electrostatic changes) can modify orbital level splittings and state mixing, thus may renormalize effective spin-Hamiltonian parameters and alter spin-dependent relaxation pathways that govern optical spin polarization and readout~\cite{udvarhelyi_planar_2023,lyu_strain_2022}. This environmental sensitivity may explain the disappearance of the ODMR signal after thermal cycling.

To further probe the underlying spin dynamics, pulsed pump--probe recovery measurements were performed (see Fig.\ref{appendix magnetic part}(b) for the pulse sequence and corresponding time trace). In this protocol, an optical pulse initializes the system, followed by a variable waiting time $\tau$, after which a readout pulse probes the recovered fluorescence. During the readout pulse, the fluorescence decays exponentially as optical excitation transfers the population to a darker, long-lived state. The amplitude of this decay depends on $\tau$, and its recovery follows a single-exponential behavior from which a characteristic relaxation time $T_1$ is extracted. The contrast refers to the relative difference between the early-time and steady-state fluorescence within each readout pulse. The extracted $T_1$ therefore represents an optically conditioned population recovery time that reflects spin relaxation under the given initialization and readout conditions.

Under off-resonant excitation at $77~\mathrm{K}$ and in the presence of the magnetic field, the recovery dynamics yield $T_1 = (5.6 \pm 0.8)\,\mathrm{ms}$ for a magnet orientation of $50^\circ$ with a fluorescence contrast of approximately $18\%$, and $T_1 = (5.1 \pm 1.3)\,\mathrm{ms}$ for $140^\circ$ with a contrast of approximately $16\%$ [Fig.~\ref{spin part}(c)]. When the magnetic field is removed, the recovery becomes faster, with $T_1 = (1.2 \pm 0.5)\,\mathrm{ms}$, and the contrast reduces to $\sim 11\%$. The slightly higher contrast at $50^\circ$ is consistent with the enhanced initialization fidelity, supporting a reduced spin-mixing rate for this field orientation. The larger transient contrast observed at $50^\circ$, together with the reduced steady-state photoluminescence intensity, indicates that this field orientation enhances population transfer into the metastable shelving state. In contrast, the higher PL and reduced contrast at $140^\circ$ suggest suppressed branching into the dark configuration. These observations are consistent with magnetic-field-controlled spin mixing that modulates the coupling between bright and dark pathways~\cite{clua-provost_spin-dependent_2024}.

The reduction of $T_1$ in the absence of the magnetic field indicates that the relaxation dynamics are sensitive to Zeeman splitting. At a finite field, the degeneracy of spin sublevels is lifted, and spin-mixing pathways are modified, which can suppress relaxation channels driven by transverse magnetic noise or cross-relaxation with nearby spin impurities. Consistently, we also observe a reduced fluorescence contrast at zero field ($\sim 11\%$ compared to $\sim 16$--$18\%$ with field). This suggests that, in the absence of a magnetic field, the optically prepared population imbalance partially relaxes already during the waiting time $\tau$, leading to a smaller transient signal during readout. Similar magnetic-field-dependent longitudinal relaxation behavior has been reported for negatively-charged Nitrogen Vacancy center (NV) in diamond, where cross-relaxation processes are strongest near degeneracies and can be tuned by applying a magnetic field~\cite{PhysRevLett.108.197601}.

Under resonant excitation, a longer recovery time of $T_1 = (9.9 \pm 1.3)\,\mathrm{ms}$ with a contrast of $\sim 28\%$ is observed in the presence of the magnetic field [Fig.~\ref{spin part}(d)]. When the field is removed, the relaxation time decreases to $T_1 = (7.2 \pm 2.1)\,\mathrm{ms}$ and the fluorescence contrast is strongly reduced to approximately $5.5\%$. Here, both the relaxation time and the contrast are strongly field-dependent under resonant excitation. This indicates that resonant optical cycling is more spin-selective and therefore more sensitive to magnetic-field-induced modifications of spin mixing.

The higher fluorescence contrast observed under resonant excitation in the presence of the magnetic field, compared to off-resonant excitation, reflects enhanced spin selectivity of the resonant optical transition. Resonant excitation addresses a specific transition within the excited-state manifold, strengthening spin-dependent branching and optical pumping into particular spin sublevels. In contrast, off-resonant excitation can populate a broader set of excited states and engage additional relaxation and charge-conversion pathways~\cite{cardoso_barbosa_impact_2023,doherty_nitrogen-vacancy_2013}, which can reduce the effective spin selectivity of the optical cycle. A similar enhancement of spin readout contrast under resonant excitation compared to nonresonant excitation has been reported for NV centers in diamond under cryogenic conditions~\cite{monge_resonant_2023}. Additionally, a recent study on hBN demonstrates that the dependence of ODMR contrast on excitation wavelength arises from different coupling strengths between optically excited states and optically inactive metastable states, further underscoring the role of excitation-selective spin dynamics~\cite{zhigulin_multi-wavelength_2026}.

Interestingly, at zero magnetic field, the fluorescence contrast under off-resonant excitation exceeds that obtained under resonant excitation. This indicates that the mechanisms generating the contrast differ between the two excitation schemes. Under off-resonant excitation, the contrast primarily reflects efficient optical pumping into a long-lived metastable state during the readout pulse, a process that is relatively insensitive to spin degeneracy. In contrast, resonant excitation relies more strongly on spin-selective optical cycling between well-defined sublevels; enhanced spin mixing at zero field reduces this selectivity and diminishes the differential fluorescence signal.

The transient decay observed during the readout pulse evidences a long-lived metastable shelving state. Its pronounced magnetic-field dependence in both recovery time and fluorescence contrast indicates a spin-dependent origin. These results establish a connection between spin-dependent shelving and the emission dynamics discussed above. The observed repumping behavior under weak blue illumination is consistent with depopulation of such shelving states, while the spectral diffusion remains dominated by charge fluctuations. Together, this supports a picture in which spin-dependent shelving governs emission intermittency, whereas environmental charge dynamics determine spectral stability.

\section{Conclusion}

In summary, we investigated the spectral stability and spin-dependent relaxation dynamics of a single, mechanically isolated quantum emitter in hexagonal boron nitride integrated on a coplanar waveguide. 

The emitter exhibits a high resonant saturation count rate of $I_\infty = 12.5 \pm 1.5~\mathrm{Mc/s}$, confirming its exceptional optical brightness and suitability for quantum photonics applications. Combined with a moderate 
Debye--Waller factor and a nanosecond excited-state lifetime, this photon flux places the defect among the brightest hBN emitters reported to date. The measured spectral-diffusion rate remains sufficiently slow to enable future coherent optical-control experiments. At an excitation power of $20~\mu\mathrm{W}$, the dominant transition exhibits a spectral-diffusion rate of approximately $80~\mathrm{kHz}$, corresponding to a characteristic timescale of $\sim 12.5~\mu\mathrm{s}$, much longer than the optical lifetime. At detected count rates above $10~\mathrm{Mc/s}$, more than 100 photons can 
be collected within this time window, providing favorable photon statistics for resolving resonant optical dynamics.

High-resolution resonant spectroscopy reveals two closely spaced ZPL transitions originating from the same defect complex. Within the DAP framework, these two lines are interpreted as distinct recombination pathways with different sensitivities to the local electrostatic environment. Such a DAP-like character could also be advantageous for future coupled-emitter devices, since the associated electric dipole moment may allow emitters to interact over larger distances and reduce the need for near-atomic defect placement~\cite{bilgin_donor-acceptor_2024}. A detailed analysis of their spectral dynamics shows that the dominant transition exhibits comparatively weak temperature dependence and moderate spectral diffusion, whereas the non-dominant transition is strongly influenced by thermally activated fluctuations. We further show that weak off-resonant blue illumination significantly increases the emission duty cycle without proportionally enhancing spectral diffusion rates at low temperature. This behavior indicates that the blue laser primarily acts as a repumping mechanism, restoring population from long-lived shelving states into the radiative manifold rather than modifying the charge environment responsible for spectral wandering which provides a practical route to increase optical cyclicity and resonant photon collection, thereby reducing the integration time required for future coherent optical-driving measurements.

Magnetic-field-dependent photoluminescence, optically detected magnetic resonance, and pulsed pump--probe recovery measurements reveal millisecond-scale relaxation dynamics and excitation-dependent spin selectivity. The strong magnetic-field dependence of the relaxation time and fluorescence contrast indicates that the metastable shelving state involved in the optical cycle possesses a spin character. The observation of a pump--probe contrast reaching approximately $30\%$ under resonant excitation is a useful figure of merit, since optical contrast directly influences spin-readout fidelity and magnetic-field sensitivity in defect-based quantum sensors. The ZPL wavelength of $585\,\mathrm{nm}$, together with the observation of ODMR, is qualitatively consistent with previous reports suggesting that the emitter can be associated with carbon-related defects~\cite{ mendelson_identifying_2021, stern_room-temperature_2022}. In addition, theoretical studies of substitutional carbon defects, have shown that carbon-related DAP defects can support optically active and paramagnetic states with comparable PL spectra and ODMR signatures~\cite{auburger_towards_2021}. However, carbon-related hBN emitters do not exhibit a single universal ODMR frequency or ZPL wavelength, since different carbon configurations, charge states, local environments, and spin-pair mechanisms can lead to different spin states and zero-field-splitting parameters~\cite{vogl_defects_2026}. While these similarities support compatibility with a carbon-related defect, they do not allow a unique microscopic identification of the emitter studied here.

Our results clarify the interplay between pathway-dependent recombination, environmental charge noise, and spin-mediated population dynamics in hBN quantum emitters. The combination of high resonant photon flux, controllable optical cyclicity, pronounced pump--probe contrast, and spin-dependent relaxation pathways highlights the potential of such emitters for future two-dimensional spin--photon interfaces, quantum-sensing platforms, and quantum photonic devices.

\section*{Acknowledgement}
We thank the Ulm Center for Nanotechnology and Quantum Material for the metal deposition. The project was funded by the German Federal Ministry of Education and Research within the research program Quantum Systems in the project
13N16741. A.K. acknowledges the support of the Baden-Württemberg Stiftung gGmbH in Project No. BWST-ISF2022-026. A.K. acknowledges the support of IQst.

\appendix

\section{On-state probability calculation via photon-event histogram}

\begin{figure}
\hspace{0cm}{\includegraphics[scale=1]{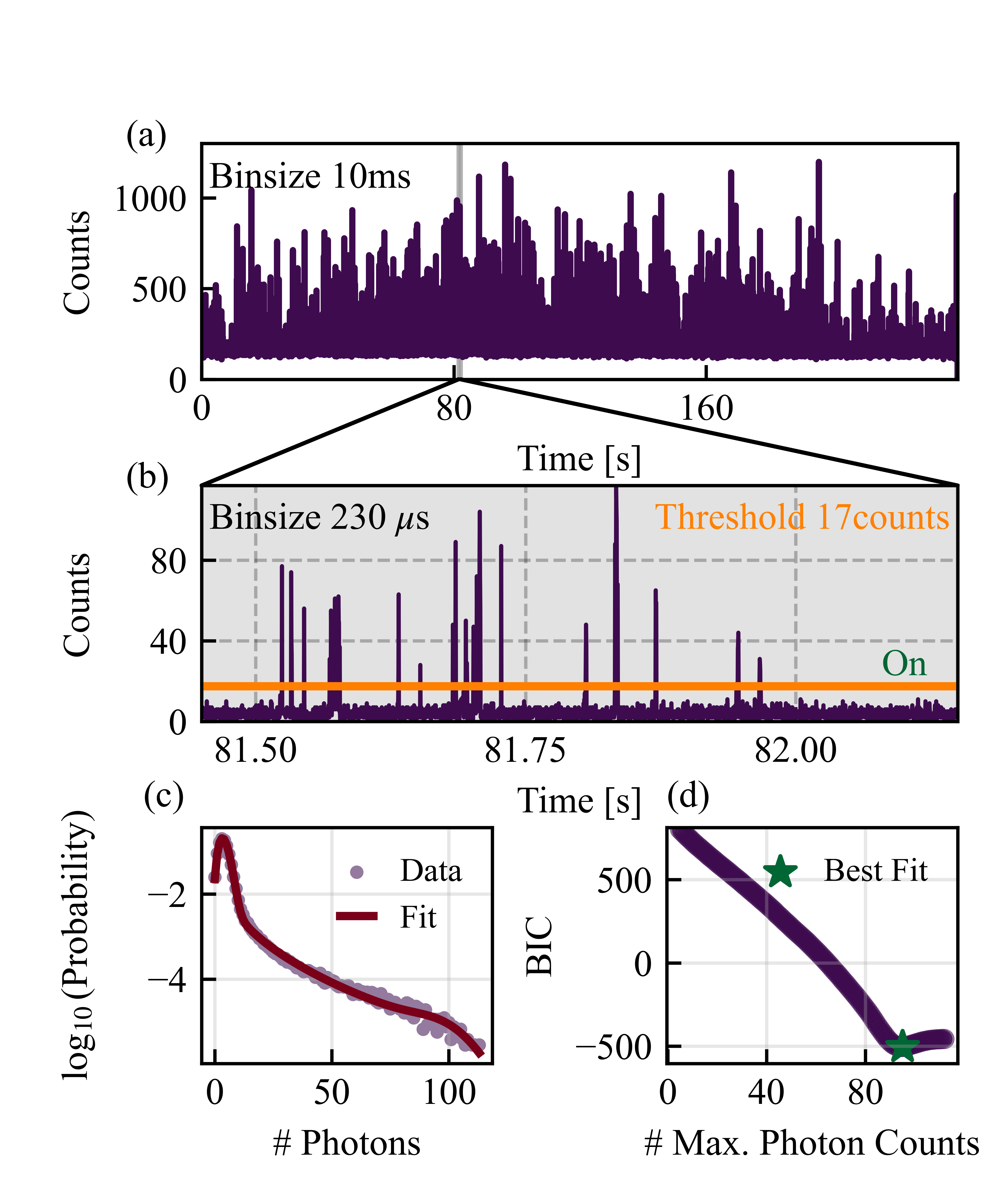}} 
\caption{Time traces and photon-count statistics of the emitter. (a) Resonant time trace of the dominant ZPL transition. (b) Zoom in on the resonant time trace from (a) with a reduced bin size. (c) Probability distribution of detected photon numbers (purple) with fit (red). (d) Bayesian information criterion for different maximum photon numbers, with the best-fit parameter marked.}
\label{fig:photon statistics}
\end{figure}

To accurately monitor the spectral diffusion dynamics, the bin size of the recorded count trace was chosen to be shorter than the characteristic spectral diffusion timescale, while remaining sufficiently long to ensure reliable detection of on-resonance events. This compromise was achieved by selecting a binning interval that undersamples background noise but still fully resolves all fluorescence events exceeding the threshold, as illustrated in Fig~\ref{fig:photon statistics}(a),(b).

The detected on-resonance events were histogrammed over time and normalized by the total number of events. Each dataset was subsequently linearized and fitted with a linear function to extract the average on-time, denoted as $\tau_{\mathrm{on}}$. From these fits, the off-diffusion rate $\gamma_{\mathrm{off}} = 1/\tau_{\mathrm{on}}$ characterizes the rate at which the emitter leaves resonance, while the on-diffusion rate $\gamma_{\mathrm{on}}$ describes the rate at which the emitter returns to resonance.

To determine the most accurate on-resonance fraction of the emitter, it is insufficient to simply count events above a threshold, as shown in Fig.~\ref{fig:photon statistics} (b), since this approach introduces bias through the threshold. Instead, the time trace is analyzed via a photon-event histogram, as shown in Fig.~\ref{fig:photon statistics} (c).

The histogram exhibits two overlapping distributions. The larger and narrower distribution is attributed to the background, while the broader distribution originates from the spectrally diffusing emitter. Spectral diffusion leads to a broadening of the estimated Poissonian photon-number distribution of the emitter. The total distribution is modeled as a weighted series of Poisson distributions $\mathbf{Po}$,
\begin{align}
P(\lambda) &= (1-p_\mathrm{e}) \mathbf{Po}(\lambda_\mathrm{b})
\nonumber \\
&\quad + p_\mathrm{e}\int_{0}^{\lambda_\mathrm{max}}
\frac{p(\lambda^\prime)e^{-\gamma \lambda^\prime}}{N}
\mathbf{Po}(\lambda^\prime)\,\mathrm{d}\lambda^\prime .
\end{align}

\begin{figure}[h]
\hspace{0cm}{\includegraphics[scale=1]{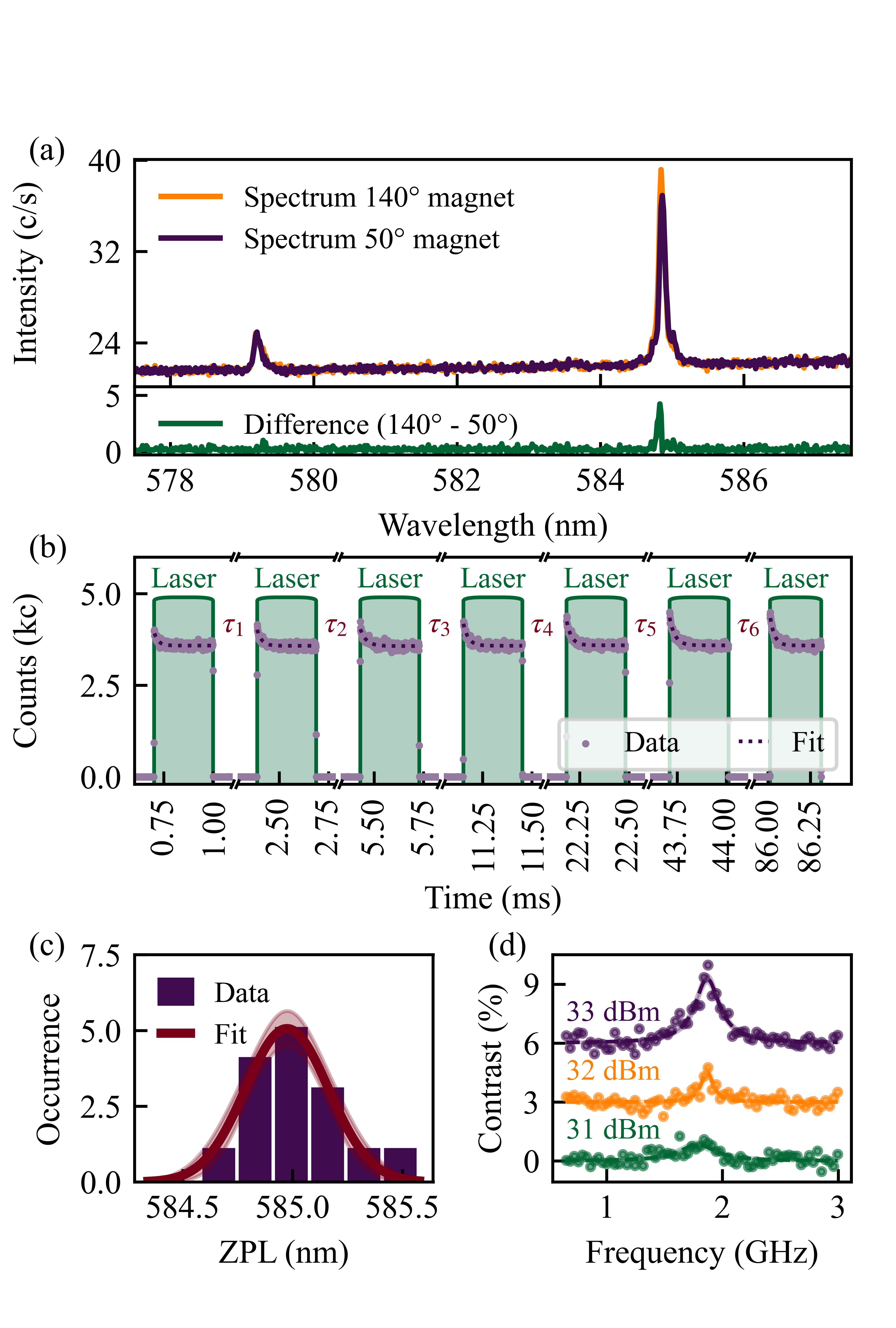}} 
\caption{Spin and spectral properties of the emitter. (a) Photoluminescence spectra at magnetic-field orientations of $50^\circ$ and $140^\circ$. 
(b) Pump--probe fluorescence time trace under off-resonant excitation (240~$\mu$W) with increasing delay $\tau_i$ from 700~ns to 42~$\mu$s. 
(c) Histogram of the ZPL center wavelength at 77~K over multiple thermal cycles, fitted with a Gaussian distribution with a $2\sigma$ value of $0.38$~nm. 
(d) ODMR spectra measured at three microwave powers showing reduced contrast with decreasing power.}
\label{appendix magnetic part}
\end{figure}

Here, $1-p_\mathrm{e}$ and $p_\mathrm{e}$ denote the probabilities of detecting photons originating from the background and the emitter, respectively. The mean photon number of the background is $\lambda_\mathrm{b}$, while the emitter's mean photon number $\lambda^\prime$ varies due to detuning induced by spectral diffusion. The detuning distribution is described by the weighting function $p(\lambda^\prime)$, which represents the probability of observing the emitter at a given detuning. The factor $e^{-\gamma\lambda^\prime}$ accounts for the pumping dynamics occurring on the time scale of the time-bin width. The normalization constant $N$ ensures that the fraction is properly normalized. The upper integration limit $\lambda_\mathrm{max}$ corresponds to the mean photon number of the emitter on resonance.

The model contains four free parameters, $p_\mathrm{e}$, $\lambda_\mathrm{b}$, $\lambda_\mathrm{max}$, and $\gamma$, and is therefore well suited to fitting despite the presence of the integral. In practice, the integral is discretized into a sum. The parameters $p_\mathrm{e}$, $\lambda_\mathrm{b}$, and $\gamma$ are fitted for different fixed values of $\lambda_\mathrm{max}$. The optimal value of $\lambda_\mathrm{max}$ is determined by minimizing the Bayesian information criterion (BIC), as shown in Fig.~\ref{fig:photon statistics} (c) and (d).

\section{Sample Geometry, Magnetic-Field-Dependent PL Spectra and pulsed scheme}

\begin{figure}
\hspace{0cm}{\includegraphics[scale=1]{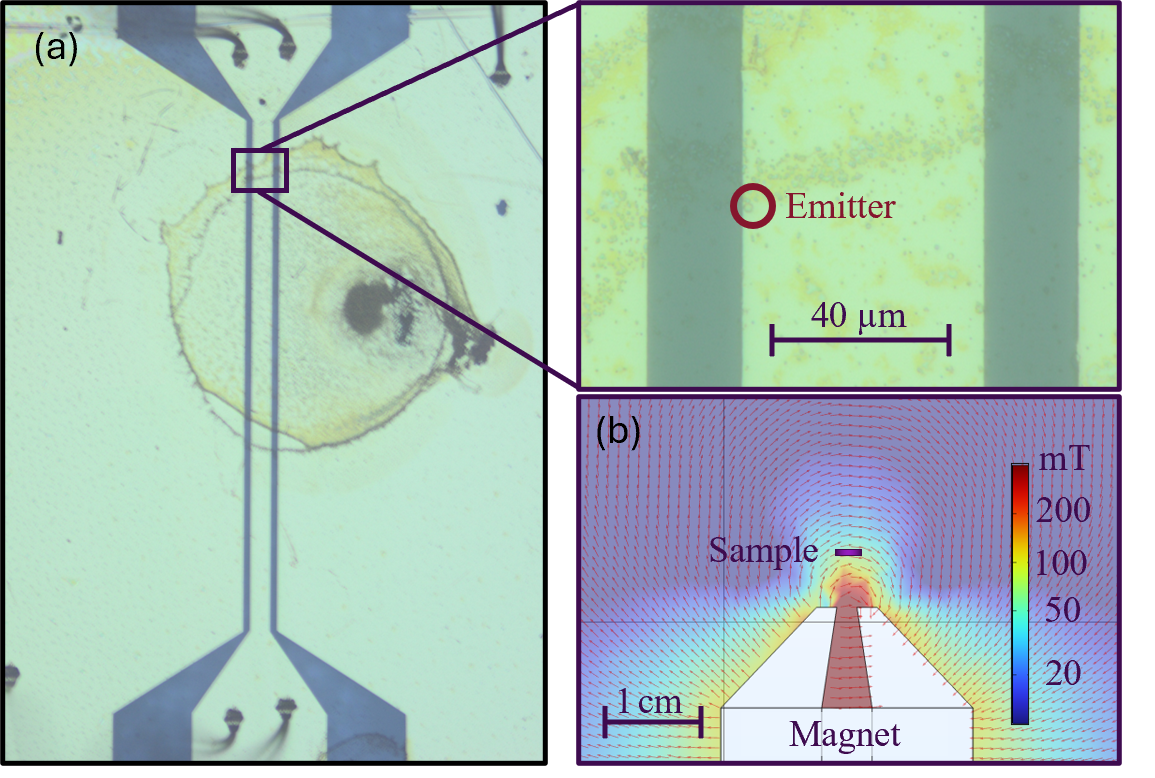}} 
\caption{(a) Optical microscope image of the tapered coplanar waveguide with spin coated hBN solutuion used in this work. 
The gold waveguide, fabricated on a sapphire substrate, is wire-bonded to a PCB for microwave delivery. (b) Magnetostatic simulation of the magnetic-field distribution generated by the horseshoe permanent magnet near the sample position.
}
\label{fig:coplanar}
\end{figure}

To verify that the observed angular modulation of the photoluminescence signal originates from the ZPL emission itself rather than from background or neighboring spectral features, full PL spectra were recorded for two representative magnetic-field orientations corresponding to the minimum ($50^\circ$) and maximum ($140^\circ$) of the oscillation shown in Fig.~\ref{spin part}(a). As shown in Fig.~\ref{appendix magnetic part}, the ZPL intensity is clearly reduced at $50^\circ$ compared to $140^\circ$, while a nearby reference emission line at 579.5~nm remains unchanged for both magnetic-field orientations. This confirms that the observed sinusoidal modulation arises from the ZPL transition and is not due to variations in background emission, collection efficiency, or optical misalignment.

Figure~\ref{appendix magnetic part}(b) presents the pump--probe fluorescence time trace under off-resonant excitation using 240~$\mu$W laser pulses, where the delay $\tau_i$ between successive pulses is increased from 700~ns up to 42~$\mu$s. The evolution of the fluorescence recovery with increasing delay illustrates the characteristic relaxation dynamics. From exponential fits to the fluorescence decay during each laser pulse, spin-pumping times of $(27 \pm 4)\mu\mathrm{s}$ and $(30 \pm 2)\mu\mathrm{s}$ were obtained in the presence and absence of the magnetic field, respectively. Similar pump--probe measurements for resonant excitation yield pumping times of $(326 \pm 85)\mu\mathrm{s}$ and $(740 \pm 170)\mu\mathrm{s}$ with and without the magnetic field, respectively. This magnetic-field dependence provides further evidence for the involvement of spin-selective processes and constitutes an additional indication of an underlying spin degree of freedom associated with the emitter.

Figure~\ref{appendix magnetic part}(d) shows the ODMR spectra measured at three different microwave powers. The ODMR contrast decreases from approximately 3\% to 1.5\% as the microwave power is reduced, consistent with a power-dependent driving efficiency of the spin transition. Figure~\ref{fig:coplanar} shows an optical microscope image of the coplanar waveguide used for the ODMR measurements with the studied emitter located on the signal line. The figure also includes a magnetostatic simulation of the horseshoe permanent magnet, illustrating the magnetic-field distribution near the sample position.



\bibliography{biblo}

@article{Mattheyses2005,
  author       = {Mattheyses, Alexa L. and Axelrod, Daniel},
  title        = {Fluorescence emission patterns near glass and metal‐coated surfaces investigated with back focal plane imaging},
  journal      = {Journal of Biomedical Optics},
  year         = {2005},
  volume       = {10},
  number       = {5},
  pages        = {054007},
  doi          = {10.1117/1.2052867},
}

@article{Nikolay:19,
author = {Niko Nikolay and Noah Mendelson and Ersan \"{O}zelci and Bernd Sontheimer and Florian B\"{o}hm and G\"{u}nter Kewes and Milos Toth and Igor Aharonovich and Oliver Benson},
journal = {Optica},
number = {8},
pages = {1084--1088},
publisher = {Optica Publishing Group},
title = {Direct measurement of quantum efficiency of single-photon emitters in hexagonal boron nitride},
volume = {6},
month = {Aug},
year = {2019},
url = {https://opg.optica.org/optica/abstract.cfm?URI=optica-6-8-1084},
doi = {10.1364/OPTICA.6.001084},
}

@Article{Castelletto2020,
author="Stefania Castelletto and Faraz A. Inam and Shin-ichiro Sato and Alberto Boretti",
title="Hexagonal boron nitride: a review of the emerging material platform for single-photon sources and the spin–photon interface",
journal="Beilstein Journal of Nanotechnology",
year="2020",
volume="11",
pages="740-769",
issn="2190-4286",
doi="10.3762/bjnano.11.61",
copyright="Castelletto et al.; licensee Beilstein-Institut",
publisher="Beilstein-Institut",
URL="https://doi.org/10.3762/bjnano.11.61",
}

@article{hoese_mechanical_2020,
	title = {Mechanical decoupling of quantum emitters in hexagonal boron nitride from low-energy phonon modes},
	volume = {6},
	url = {https://www.science.org/doi/full/10.1126/sciadv.aba6038},
	doi = {10.1126/sciadv.aba6038},
	pages = {eaba6038},
	number = {40},
	journal = {Science Advances},
	publisher = {American Association for the Advancement of Science},
	author = {Hoese, Michael and Reddy, Prithvi and Dietrich, Andreas and Koch, Michael K. and Fehler, Konstantin G. and Doherty, Marcus W. and Kubanek, Alexander},
	urldate = {2025-02-19},
	year = {2020-09-30},
}

@article{mejia_dynamic_2025,
	title = {Dynamic Interplay of Nonlocal Recombination Pathways in Quantum Emitters in Hexagonal Boron Nitride},
	volume = {129},
	url = {https://doi.org/10.1021/acs.jpcc.4c07147},
	doi = {10.1021/acs.jpcc.4c07147},
	pages = {2044--2053},
	number = {4},
	journal = {The Journal of Physical Chemistry C},
	author = {Mejia, Enrique A. and Woods, John M. and Adhikari, Ashok and Singh, Charanjot and Taniguchi, Takashi and Watanabe, Kenji and Bisogni, Valentina and Sofer, Zdeněk and Pelliciari, Jonathan and Grosso, Gabriele},
	year = {2025},
	note = {\_eprint: https://doi.org/10.1021/acs.jpcc.4c07147},
}

@article{pelliciari_elementary_2024,
	title = {Elementary excitations of single-photon emitters in hexagonal boron nitride},
	volume = {23},
	issn = {1476-4660},
	url = {https://doi.org/10.1038/s41563-024-01866-4},
	doi = {10.1038/s41563-024-01866-4},
	pages = {1230--1236},
	number = {9},
	journal= {Nature Materials},
	shortjournal = {Nature Materials},
	author = {Pelliciari, Jonathan and Mejia, Enrique and Woods, John M. and Gu, Yanhong and Li, Jiemin and Chand, Saroj B. and Fan, Shiyu and Watanabe, Kenji and Taniguchi, Takashi and Bisogni, Valentina and Grosso, Gabriele},
	year = {2024-09-01},
}

@article{PhysRevLett.108.197601,
  title = {Temperature- and Magnetic-Field-Dependent Longitudinal Spin Relaxation in Nitrogen-Vacancy Ensembles in Diamond},
  author = {Jarmola, A. and Acosta, V. M. and Jensen, K. and Chemerisov, S. and Budker, D.},
  journal = {Phys. Rev. Lett.},
  volume = {108},
  issue = {19},
  pages = {197601},
  numpages = {5},
  year = {2012},
  month = {May},
  publisher = {American Physical Society},
  doi = {10.1103/PhysRevLett.108.197601},
  url = {https://link.aps.org/doi/10.1103/PhysRevLett.108.197601}
}

@article{exarhos_magnetic-field-dependent_2019,
	title = {Magnetic-field-dependent quantum emission in hexagonal boron nitride at room temperature},
	volume = {10},
	rights = {2019 The Author(s)},
	issn = {2041-1723},
	url = {https://www.nature.com/articles/s41467-018-08185-8},
	doi = {10.1038/s41467-018-08185-8},
	pages = {222},
	number = {1},
	journal = {Nature Communications},
	shortjournal = {Nat Commun},
	publisher = {Nature Publishing Group},
	author = {Exarhos, Annemarie L. and Hopper, David A. and Patel, Raj N. and Doherty, Marcus W. and Bassett, Lee C.},
	urldate = {2026-02-11},
	year = {2019-01-15},
	langid = {english},
}

@article{monge_resonant_2023,
	title = {Resonant Versus Nonresonant Spin Readout of a Nitrogen-Vacancy Center in Diamond under Cryogenic Conditions},
	volume = {131},
	url = {https://link.aps.org/doi/10.1103/PhysRevLett.131.236901},
	doi = {10.1103/PhysRevLett.131.236901},
	pages = {236901},
	number = {23},
	journal = {Physical Review Letters},
	shortjournal = {Phys. Rev. Lett.},
	publisher = {American Physical Society},
	author = {Monge, Richard and Delord, Tom and Thiering, Gergo and Gali, Adam and Meriles, Carlos A.},
	urldate = {2026-02-11},
	year = {2023-12-05},
}

@article{cardoso_barbosa_impact_2023,
	title = {Impact of charge conversion on {NV}-center relaxometry},
	volume = {108},
	url = {https://link.aps.org/doi/10.1103/PhysRevB.108.075411},
	doi = {10.1103/PhysRevB.108.075411},
	pages = {075411},
	number = {7},
	journal = {Physical Review B},
	shortjournal = {Phys. Rev. B},
	publisher = {American Physical Society},
	author = {Cardoso Barbosa, Isabel and Gutsche, Jonas and Widera, Artur},
	urldate = {2026-02-11},
	year = {2023-08-14},
}

@article{doherty_nitrogen-vacancy_2013,
	title = {The nitrogen-vacancy colour centre in diamond},
	volume = {528},
	issn = {0370-1573},
	url = {https://www.sciencedirect.com/science/article/pii/S0370157313000562},
	doi = {10.1016/j.physrep.2013.02.001},
	series = {The nitrogen-vacancy colour centre in diamond},
	pages = {1--45},
	number = {1},
	journal = {Physics Reports},
	shortjournal = {Physics Reports},
	author = {Doherty, Marcus W. and Manson, Neil B. and Delaney, Paul and Jelezko, Fedor and Wrachtrup, Jörg and Hollenberg, Lloyd C. L.},
	urldate = {2026-02-11},
	year = {2013-07-01},
}

@article{clua-provost_spin-dependent_2024,
	title = {Spin-dependent photodynamics of boron-vacancy centers in hexagonal boron nitride},
	volume = {110},
	url = {https://link.aps.org/doi/10.1103/PhysRevB.110.014104},
	doi = {10.1103/PhysRevB.110.014104},
	pages = {014104},
	number = {1},
	journal = {Physical Review B},
	shortjournal = {Phys. Rev. B},
	publisher = {American Physical Society},
	author = {Clua-Provost, T. and Mu, Z. and Durand, A. and Schrader, C. and Happacher, J. and Bocquel, J. and Maletinsky, P. and Fraunié, J. and Marie, X. and Robert, C. and Seine, G. and Janzen, E. and Edgar, J. H. and Gil, B. and Cassabois, G. and Jacques, V.},
	urldate = {2026-02-13},
	year = {2024-07-11},
}

@article{stern_quantum_2024,
	title = {A quantum coherent spin in hexagonal boron nitride at ambient conditions},
	volume = {23},
	issn = {1476-1122, 1476-4660},
	url = {https://www.nature.com/articles/s41563-024-01887-z},
	doi = {10.1038/s41563-024-01887-z},
	pages = {1379--1385},
	number = {10},
	journal = {Nature Materials},
	shortjournal = {Nat. Mater.},
	author = {Stern, Hannah L. and M. Gilardoni, Carmem and Gu, Qiushi and Eizagirre Barker, Simone and Powell, Oliver F. J. and Deng, Xiaoxi and Fraser, Stephanie A. and Follet, Louis and Li, Chi and Ramsay, Andrew J. and Tan, Hark Hoe and Aharonovich, Igor and Atature, Mete},
	urldate = {2025-10-01},
	year = {2024-10},
	langid = {english},
}

@article{whitefield_narrowband_2026,
	title = {Narrowband quantum emitters in hexagonal boron nitride with optically addressable spins},
	rights = {2026 The Author(s), under exclusive licence to Springer Nature Limited},
	issn = {1476-4660},
	url = {https://www.nature.com/articles/s41563-025-02458-6},
	doi = {10.1038/s41563-025-02458-6},
	pages = {1--8},
	journal = {Nature Materials},
	shortjournal = {Nat. Mater.},
	publisher = {Nature Publishing Group},
	author = {Whitefield, Benjamin and Zeng, Helen Zhi Jie and Liddle-Wesolowski, James and Robertson, Islay O. and Ganyecz, adam and Ivady, Viktor and Watanabe, Kenji and Taniguchi, Takashi and Toth, Milos and Tetienne, Jean-Philippe and Aharonovich, Igor and Kianinia, Mehran},
	urldate = {2026-02-16},
	year = {2026-01-27},
	langid = {english},
}

@article{gottscholl_room_2021,
	title = {Room temperature coherent control of spin defects in hexagonal boron nitride},
	volume = {7},
	url = {https://www.science.org/doi/10.1126/sciadv.abf3630},
	doi = {10.1126/sciadv.abf3630},
	pages = {eabf3630},
	number = {14},
	journal = {Science Advances},
	publisher = {American Association for the Advancement of Science},
	author = {Gottscholl, Andreas and Diez, Matthias and Soltamov, Victor and Kasper, Christian and Sperlich, Andreas and Kianinia, Mehran and Bradac, Carlo and Aharonovich, Igor and Dyakonov, Vladimir},
	urldate = {2026-02-20},
	year = {2021-04-02},
}

@article{grosso_tunable_2017,
	title = {Tunable and high-purity room temperature single-photon emission from atomic defects in hexagonal boron nitride},
	volume = {8},
	rights = {2017 The Author(s)},
	issn = {2041-1723},
	url = {https://www.nature.com/articles/s41467-017-00810-2},
	doi = {10.1038/s41467-017-00810-2},
	pages = {705},
	number = {1},
	journal = {Nature Communications},
	shortjournal = {Nat Commun},
	publisher = {Nature Publishing Group},
	author = {Grosso, Gabriele and Moon, Hyowon and Lienhard, Benjamin and Ali, Sajid and Efetov, Dmitri K. and Furchi, Marco M. and Jarillo-Herrero, Pablo and Ford, Michael J. and Aharonovich, Igor and Englund, Dirk},
	urldate = {2026-02-20},
	year = {2017-09-26},
	langid = {english},
}

@article{noh_stark_2018,
	title = {Stark Tuning of Single-Photon Emitters in Hexagonal Boron Nitride},
	volume = {18},
	issn = {1530-6984},
	url = {https://doi.org/10.1021/acs.nanolett.8b01030},
	doi = {10.1021/acs.nanolett.8b01030},
	pages = {4710--4715},
	number = {8},
	journal = {Nano Letters},
	shortjournal = {Nano Lett.},
	publisher = {American Chemical Society},
	author = {Noh, Gichang and Choi, Daebok and Kim, Jin-Hun and Im, Dong-Gil and Kim, Yoon-Ho and Seo, Hosung and Lee, Jieun},
	urldate = {2026-02-20},
	year= {2018-08-08},
}

@article{white_phonon_2021,
	title = {Phonon dephasing and spectral diffusion of quantum emitters in hexagonal boron nitride},
	volume = {8},
	rights = {© 2021 Optical Society of America},
	issn = {2334-2536},
	url = {https://opg.optica.org/optica/abstract.cfm?uri=optica-8-9-1153},
	doi = {10.1364/OPTICA.431262},
	pages = {1153--1158},
	number = {9},
	journal = {Optica},
	shortjournal = {Optica, {OPTICA}},
	publisher = {Optica Publishing Group},
	author = {White, Simon and Stewart, Connor and Solntsev, Alexander S. and Li, Chi and Toth, Milos and Kianinia, Mehran and Aharonovich, Igor},
	urldate = {2026-02-20},
	year = {2021-09-20},
}

@article{udvarhelyi_planar_2023,
	title = {A planar defect spin sensor in a two-dimensional material susceptible to strain and electric fields},
	volume = {9},
	rights = {2023 The Author(s)},
	issn = {2057-3960},
	url = {https://www.nature.com/articles/s41524-023-01111-7},
	doi = {10.1038/s41524-023-01111-7},
	pages = {150},
	number = {1},
	journal = {npj Computational Materials},
	shortjournal = {npj Comput Mater},
	publisher = {Nature Publishing Group},
	author = {Udvarhelyi, Péter and Clua-Provost, Tristan and Durand, Alrik and Li, Jiahan and Edgar, James H. and Gil, Bernard and Cassabois, Guillaume and Jacques, Vincent and Gali, Adam},
	urldate = {2026-02-20},
	year = {2023-08-22},
	langid = {english},
}

@article{lyu_strain_2022,
	title = {Strain Quantum Sensing with Spin Defects in Hexagonal Boron Nitride},
	volume = {22},
	issn = {1530-6984},
	url = {https://doi.org/10.1021/acs.nanolett.2c01722},
	doi = {10.1021/acs.nanolett.2c01722},
	pages = {6553--6559},
	number = {16},
	journal = {Nano Letters},
	shortjournal = {Nano Lett.},
	publisher = {American Chemical Society},
	author = {Lyu, Xiaodan and Tan, Qinghai and Wu, Lishu and Zhang, Chusheng and Zhang, Zhaowei and Mu, Zhao and Zúñiga-Pérez, Jesús and Cai, Hongbing and Gao, Weibo},
	urldate = {2026-02-20},
	year = {2022-08-24},
}

@article{bourrellier_bright_2016,
	title = {Bright {UV} Single Photon Emission at Point Defects in h-{BN}},
	volume = {16},
	issn = {1530-6984},
	url = {https://doi.org/10.1021/acs.nanolett.6b01368},
	doi = {10.1021/acs.nanolett.6b01368},
	pages = {4317--4321},
	number = {7},
	journal = {Nano Letters},
	shortjournal = {Nano Lett.},
	publisher = {American Chemical Society},
	author = {Bourrellier, Romain and Meuret, Sophie and Tararan, Anna and Stéphan, Odile and Kociak, Mathieu and Tizei, Luiz H. G. and Zobelli, Alberto},
	urldate = {2026-02-24},
	year = {2016-07-13},
}

@article{dietrich_observation_2018,
	title = {Observation of Fourier transform limited lines in hexagonal boron nitride},
	volume = {98},
	url = {https://link.aps.org/doi/10.1103/PhysRevB.98.081414},
	doi = {10.1103/PhysRevB.98.081414},
	pages = {081414},
	number = {8},
	journal = {Physical Review B},
	shortjournal = {Phys. Rev. B},
	publisher = {American Physical Society},
	author = {Dietrich, A. and Bürk, M. and Steiger, E. S. and Antoniuk, L. and Tran, T. T. and Nguyen, M. and Aharonovich, I. and Jelezko, F. and Kubanek, A.},
	urldate = {2026-02-24},
	year = {2018-08-31},
}

@article{boll_photophysics_2020,
	title = {Photophysics of quantum emitters in hexagonal boron-nitride nano-flakes},
	volume = {28},
	rights = {© 2020 Optical Society of America},
	issn = {1094-4087},
	url = {https://opg.optica.org/oe/abstract.cfm?uri=oe-28-5-7475},
	doi = {10.1364/OE.386629},
	pages = {7475--7487},
	number = {5},
	journal = {Optics Express},
	shortjournal = {Opt. Express, {OE}},
	publisher = {Optica Publishing Group},
	author = {Boll, Mads K. and Radko, Ilya P. and Huck, Alexander and Andersen, Ulrik L.},
	urldate = {2026-02-24},
	year = {2020-03-02},
}

@article{tan_donoracceptor_2022,
	title = {Donor–Acceptor Pair Quantum Emitters in Hexagonal Boron Nitride},
	volume = {22},
	issn = {1530-6984},
	url = {https://doi.org/10.1021/acs.nanolett.1c04647},
	doi = {10.1021/acs.nanolett.1c04647},
	pages = {1331--1337},
	number = {3},
	journal = {Nano Letters},
	shortjournal = {Nano Lett.},
	publisher = {American Chemical Society},
	author = {Tan, Qinghai and Lai, Jia-Min and Liu, Xue-Lu and Guo, Dan and Xue, Yongzhou and Dou, Xiuming and Sun, Bao-Quan and Deng, Hui-Xiong and Tan, Ping-Heng and Aharonovich, Igor and Gao, Weibo and Zhang, Jun},
	urldate = {2026-02-25},
	year = {2022-02-09},
}

@article{museur_defect-related_2008,
	title = {Defect-related photoluminescence of hexagonal boron nitride},
	volume = {78},
	url = {https://link.aps.org/doi/10.1103/PhysRevB.78.155204},
	doi = {10.1103/PhysRevB.78.155204},
	pages = {155204},
	number = {15},
	journal = {Physical Review B},
	shortjournal = {Phys. Rev. B},
	publisher = {American Physical Society},
	author = {Museur, Luc and Feldbach, Eduard and Kanaev, Andrei},
	urldate = {2026-02-25},
	year = {2008-10-16},
}

@article{li_quantum_2025,
	title = {Quantum emission from coupled spin pairs in hexagonal boron nitride},
	volume = {16},
	rights = {2025 The Author(s)},
	issn = {2041-1723},
	url = {https://www.nature.com/articles/s41467-025-61388-8},
	doi = {10.1038/s41467-025-61388-8},
	pages = {5842},
	number = {1},
	journal = {Nature Communications},
	shortjournal = {Nat Commun},
	publisher = {Nature Publishing Group},
	author = {Li, Song and Pershin, Anton and Gali, Adam},
	urldate = {2026-02-25},
	year = {2025-07-01},
	langid = {english},
}

@article{koch_probing_2024,
	title = {Probing the limits for coherent optical control of a mechanically decoupled defect center in hexagonal boron nitride},
	volume = {5},
	rights = {2024 The Author(s)},
	issn = {2662-4443},
	url = {https://www.nature.com/articles/s43246-024-00686-y},
	doi = {10.1038/s43246-024-00686-y},
	pages = {1--7},
	number = {1},
	journal = {Communications Materials},
	shortjournal = {Commun Mater},
	publisher = {Nature Publishing Group},
	author = {Koch, Michael K. and Bharadwaj, Vibhav and Kubanek, Alexander},
	urldate = {2025-01-15},
	year = {2024-11-03},
	langid = {english},
}

@article{dietrich_solid-state_2020,
	title = {Solid-state single photon source with Fourier transform limited lines at room temperature},
	volume = {101},
	url = {https://link.aps.org/doi/10.1103/PhysRevB.101.081401},
	doi = {10.1103/PhysRevB.101.081401},
	pages = {081401},
	number = {8},
	journal = {Physical Review B},
	shortjournal = {Phys. Rev. B},
	publisher = {American Physical Society},
	author = {Dietrich, A. and Doherty, M. W. and Aharonovich, I. and Kubanek, A.},
	urldate = {2024-11-25},
	year = {2020-02-05},
}

@article{hoese_single_2022,
	title = {Single photon randomness originating from the symmetric dipole emission pattern of quantum emitters},
	volume = {120},
	issn = {0003-6951},
	url = {https://doi.org/10.1063/5.0074946},
	doi = {10.1063/5.0074946},
	pages = {044001},
	number = {4},
	journal = {Applied Physics Letters},
	shortjournal = {Appl. Phys. Lett.},
	author = {Hoese, Michael and Koch, Michael K. and Breuning, Felix and Lettner, Niklas and Fehler, Konstantin G. and Kubanek, Alexander},
	urldate = {2026-03-20},
	year = {2022-01-24},
}

@article{gao_single_2025,
	title = {Single nuclear spin detection and control in a van der Waals material},
	volume = {643},
	rights = {2025 The Author(s)},
	issn = {1476-4687},
	url = {https://www.nature.com/articles/s41586-025-09258-7},
	doi = {10.1038/s41586-025-09258-7},
	pages = {943--949},
	number = {8073},
	journal = {Nature},
	publisher = {Nature Publishing Group},
	author = {Gao, Xingyu and Vaidya, Sumukh and Li, Kejun and Ge, Zhun and Dikshit, Saakshi and Zhang, Shimin and Ju, Peng and Shen, Kunhong and Jin, Yuanbin and Ping, Yuan and Li, Tongcang},
	urldate = {2025-10-01},
	year = {2025-07},
	langid = {english},
}

@article{barry_sensitivity_2020,
	title = {Sensitivity optimization for {NV}-diamond magnetometry},
	volume = {92},
	url = {https://link.aps.org/doi/10.1103/RevModPhys.92.015004},
	doi = {10.1103/RevModPhys.92.015004},
	pages = {015004},
	number = {1},
	journal = {Reviews of Modern Physics},
	shortjournal = {Rev. Mod. Phys.},
	publisher = {American Physical Society},
	author = {Barry, John F. and Schloss, Jennifer M. and Bauch, Erik and Turner, Matthew J. and Hart, Connor A. and Pham, Linh M. and Walsworth, Ronald L.},
	urldate = {2024-02-08},
	year = {2020-03-31},
}

@article{galland_two_2011,
	title = {Two types of luminescence blinking revealed by spectroelectrochemistry of single quantum dots},
	volume = {479},
	rights = {2011 Springer Nature Limited},
	issn = {1476-4687},
	url = {https://www.nature.com/articles/nature10569},
	doi = {10.1038/nature10569},
	pages = {203--207},
	number = {7372},
	journal = {Nature},
	publisher = {Nature Publishing Group},
	author = {Galland, Christophe and Ghosh, Yagnaseni and Steinbrück, Andrea and Sykora, Milan and Hollingsworth, Jennifer A. and Klimov, Victor I. and Htoon, Han},
	urldate = {2026-04-10},
	year = {2011-11},
	langid = {english},
}

@article{akbari_temperature-dependent_2021,
	title = {Temperature-dependent Spectral Emission of Hexagonal Boron Nitride Quantum Emitters on Conductive and Dielectric Substrates},
	volume = {15},
	url = {https://link.aps.org/doi/10.1103/PhysRevApplied.15.014036},
	doi = {10.1103/PhysRevApplied.15.014036},
	pages = {014036},
	number = {1},
	journal = {Physical Review Applied},
	shortjournal = {Phys. Rev. Appl.},
	publisher = {American Physical Society},
	author = {Akbari, Hamidreza and Lin, Wei-Hsiang and Vest, Benjamin and Jha, Pankaj K. and Atwater, Harry A.},
	urldate = {2025-02-19},
	year = {2021-01-20},
}

@article{chejanovsky_single-spin_2021,
	title = {Single-spin resonance in a van der Waals embedded paramagnetic defect},
	volume = {20},
	rights = {2021 The Author(s), under exclusive licence to Springer Nature Limited},
	issn = {1476-4660},
	url = {https://www.nature.com/articles/s41563-021-00979-4},
	doi = {10.1038/s41563-021-00979-4},
	pages = {1079--1084},
	number = {8},
	journal = {Nature Materials},
	shortjournal = {Nat. Mater.},
	publisher = {Nature Publishing Group},
	author = {Chejanovsky, Nathan and Mukherjee, Amlan and Geng, Jianpei and Chen, Yu-Chen and Kim, Youngwook and Denisenko, Andrej and Finkler, Amit and Taniguchi, Takashi and Watanabe, Kenji and Dasari, Durga Bhaktavatsala Rao and Auburger, Philipp and Gali, Adam and Smet, Jurgen H. and Wrachtrup, Jörg},
	urldate = {2026-02-20},
	year = {2021-08},
	langid = {english},
}

@article{auburger_towards_2021,
	title = {Towards \textit{ab initio} identification of paramagnetic substitutional carbon defects in hexagonal boron nitride acting as quantum bits},
	volume = {104},
	issn = {2469-9950, 2469-9969},
	url = {https://link.aps.org/doi/10.1103/PhysRevB.104.075410},
	doi = {10.1103/PhysRevB.104.075410},
	pages = {075410},
	number = {7},
	journal = {Physical Review B},
	shortjournal = {Phys. Rev. B},
	author = {Auburger, Philipp and Gali, Adam},
	urldate = {2025-10-01},
	year = {2021-08-06},
	langid = {english},
}

@article{bilgin_donor-acceptor_2024,
	title = {Donor-acceptor pairs in wide-bandgap semiconductors for quantum technology applications},
	volume = {10},
	rights = {2024 The Author(s)},
	issn = {2057-3960},
	url = {https://www.nature.com/articles/s41524-023-01190-6},
	doi = {10.1038/s41524-023-01190-6},
	pages = {7},
	number = {1},
	journal = {npj Computational Materials},
	shortjournal = {npj Comput Mater},
	publisher = {Nature Publishing Group},
	author = {Bilgin, Anil and Hammock, Ian N. and Estes, Jeremy and Jin, Yu and Bernien, Hannes and High, Alexander A. and Galli, Giulia},
	urldate = {2026-05-08},
	year = {2024-01-06},
	langid = {english},
}

@article{vogl_defects_2026,
	title = {Defects in hexagonal boron nitride for quantum technologies: a perspective},
	volume = {13},
	issn = {2053-1583},
	url = {https://iopscience.iop.org/article/10.1088/2053-1583/ae566e},
	doi = {10.1088/2053-1583/ae566e},
	shorttitle = {Defects in hexagonal boron nitride for quantum technologies},
	pages = {023001},
	number = {2},
	journal = {2D Materials},
	shortjournal = {2D Mater.},
	author = {Vogl, Tobias and Ivády, Viktor and Luxmoore, Isaac J and Stern, Hannah L},
	urldate = {2026-05-06},
	year = {2026-06-01},
	langid = {english},
}

@article{mendelson_identifying_2021,
	title = {Identifying carbon as the source of visible single-photon emission from hexagonal boron nitride},
	volume = {20},
	rights = {2020 The Author(s), under exclusive licence to Springer Nature Limited},
	issn = {1476-4660},
	url = {https://www.nature.com/articles/s41563-020-00850-y},
	doi = {10.1038/s41563-020-00850-y},
	pages = {321--328},
	number = {3},
	journal = {Nature Materials},
	shortjournal = {Nat. Mater.},
	publisher = {Nature Publishing Group},
	author = {Mendelson, Noah and Chugh, Dipankar and Reimers, Jeffrey R. and Cheng, Tin S. and Gottscholl, Andreas and Long, Hu and Mellor, Christopher J. and Zettl, Alex and Dyakonov, Vladimir and Beton, Peter H. and Novikov, Sergei V. and Jagadish, Chennupati and Tan, Hark Hoe and Ford, Michael J. and Toth, Milos and Bradac, Carlo and Aharonovich, Igor},
	urldate = {2026-08-13},
	year = {2021-03},
	langid = {english},
}

@article{stern_room-temperature_2022,
	title = {Room-temperature optically detected magnetic resonance of single defects in hexagonal boron nitride},
	volume = {13},
	issn = {2041-1723},
	url = {https://www.nature.com/articles/s41467-022-28169-z},
	doi = {10.1038/s41467-022-28169-z},
	pages = {618},
	number = {1},
	journal = {Nature Communications},
	shortjournal = {Nat Commun},
	author = {Stern, Hannah L. and Gu, Qiushi and Jarman, John and Eizagirre Barker, Simone and Mendelson, Noah and Chugh, Dipankar and Schott, Sam and Tan, Hoe H. and Sirringhaus, Henning and Aharonovich, Igor and Atatüre, Mete},
	urldate = {2024-11-25},
	year = {2022-02-01},
	langid = {english},
}

@article{zhigulin_multi-wavelength_2026,
	title = {Multi-wavelength spin dynamics of defects in hexagonal boron nitride},
	volume = {15},
	rights = {2026 The Author(s)},
	issn = {2047-7538},
	url = {https://www.nature.com/articles/s41377-026-02398-z},
	doi = {10.1038/s41377-026-02398-z},
	pages = {283},
	number = {1},
	journal = {Light: Science \& Applications},
	shortjournal = {Light Sci Appl},
	publisher = {Nature Publishing Group},
	author = {Zhigulin, Ivan and Sloane, Nicholas P. and Whitefield, Benjamin and Shimazaki, Konosuke and Tetienne, Jean-Philippe and Kianinia, Mehran and Aharonovich, Igor},
	urldate = {2026-08-24},
	year = {2026-06-25},
	langid = {english},
}

\end{document}